\documentstyle[epsf]{aipproc}

\tighten

\begin{document}


\newcommand{\Bd}{{\dot B}}
\newcommand{\Cd}{{\dot C}}
\newcommand{\hd}{{\dot h}}
\newcommand{\ep}{\epsilon}
\newcommand{\vp}{\varphi}
\newcommand{\al}{\alpha}
\newcommand{\be}{\begin{equation}}
\newcommand{\ee}{\end{equation}}
\newcommand{\bea}{\begin{eqnarray}}
\newcommand{\eea}{\end{eqnarray}}
\def\gapp{\mathrel{\raise.3ex\hbox{$>$}\mkern-14mu
              \lower0.6ex\hbox{$\sim$}}}
\def\gsim{\gapp}
\def\lapp{\mathrel{\raise.3ex\hbox{$<$}\mkern-14mu
              \lower0.6ex\hbox{$\sim$}}}
\def\lsim{\lapp}
\newcommand{\PSbox}[3]{\mbox{\rule{0in}{#3}\includegraphics{#1}\hspace{#2}}}
\def\Tr{\mathop{\rm Tr}\nolimits}
\def\su#1{{\rm SU}(#1)}

\title{Remarks on Inflation}

\author{
Tanmay Vachaspati
}
\address
{Department of Physics,
Case Western Reserve University,
10900 Euclid Avenue,\\
Cleveland, OH 44106-7079, USA.}

\maketitle

\begin{abstract}
It has been shown that sub-Planckian models of inflation
require initial homogeneity on super-Hubble scales under
certain commonly held assumptions. Here I remark on the possible
implications of this result for inflationary cosmology.
\end{abstract}


The observed homogeneity of the universe on superhorizon
scales can be explained if there was a period of accelerated
expansion (inflation) in the early universe that took a relatively
small homogeneous patch of space and blew it up to encompass 
a volume larger than what we observe today. 
In a recent paper \cite{VacTro98}, Mark Trodden and I addressed
the issue of how large the initial homogeneous patch has to
be. The result we obtained is that there is a lower
bound on the inflationary horizon, $H_{inf}^{-1}$, which depends
on the pre-inflationary cosmology. In particular, if the 
pre-inflationary cosmology is a Friedman-Robertson-Walker 
(FRW) universe, we must have $H_{inf}^{-1} > H_{FRW}^{-1}$. 
If we further assume that, to get inflation with Hubble constant 
$H_{inf}^{-1}$, the appropriate inflationary conditions 
(homogeneity, vacuum domination etc.)
must be satisfied over a region
of physical size $L$ which is larger than $H_{inf}^{-1}$, then
we obtain $L > H_{FRW}^{-1}$. 
Hence the conditions for inflation need to be 
satisfied on cosmological scales specified by the pre-inflationary
epoch. In the particular case of a radiation dominated 
FRW, the problem seems to be even more severe since the causal
horizon coincides with $H_{FRW}^{-1}$. Hence, it is clear
that to solve the homogeneity problem, inflationary models 
in which inflation emerges from a non-inflationary, classical
epoch, must assume large-scale homogeneity as an initial condition.

Perhaps more important than the result itself is the fact that we
have identified the conditions under which such a derivation is
possible. The key assumptions are that Einstein's equations and 
the weak energy conditions are valid, and that spacetime 
topology is trivial. (We also assume that singularities apart 
from the big bang are absent.)
If these conditions hold, one would conclude that sub-Planckian
inflation alleviates the large-scale homogeneity problem but does not
{\em solve} it\footnote{The word ``solve'' may hold different meanings
for different individuals. My view is that a solution should not
assume the result it purports to obtain.}.

Note that there are two related issues that are brought to the
forefront in the above discussion. The first is - do 
inflationary models (within the stipulated assumptions) solve the 
homogeneity problem? - and my answer to this question is in the
negative. The second is - can inflation occur, 
given that it seems to require unlikely initial conditions?
Here the answer is in the positive - inflation can indeed
occur as long as we have a suitable physical theory. 
However, one may wish to 
further consider the likelihood of having no inflation 
or the probabilities with which various kinds of inflation can 
occur, and this is where the questions become difficult.
Hopefully some of the issues involved will 
become clearer by the end of this article.


A way around the result in \cite{VacTro98} is to consider
inflationary models in which inflation starts at the Planck epoch 
(for example, chaotic inflation \cite{Linrefs}) since these do not contain a 
classical pre-inflationary epoch. Hence, at least as far as classical 
physics goes, inflation is imposed as an initial condition in these 
models. These initial conditions are sometimes justified by using
quantum cosmology -- if the wavefunction of the universe is 
``highly peaked'' around the inflationary initial conditions, one might
say that these conditions are favored. However, what if the wavefunction 
only has a small tail around the inflationary initial conditions? I do not
think that that would exclude an inflating epoch of our universe since,
it could be argued, that most of the other non-inflating universes
(where the peak of the wave-function is) would not be able to harbor 
observers such as us. Hence the argument appears to be inconclusive
at present. On the other hand, anthropic arguments are essential 
to any theory in which the creation of universes is probabilistic. 
As our understanding of astrophysics and gravity improves,
the arguments are likely to get sharpened. As a very simple 
example of possible forthcoming refinements, if we assume that life 
can only exist on planets, then an understanding of the cosmological 
conditions leading to maximal planetary formation
will help in narrowing down a measure for calculating probabilities.

Guth \cite{Guth} has given a persuasive argument for believing 
that inflation took place in the early universe in spite of any 
required unlikely initial conditions. He likens inflationary cosmology
to the evolution of life. Today we observe a rich variety of life
forms, and also a large homogeneous universe.
It is hard to imagine how all the miraculous forms
of life could have been created directly. Similarly, it is hard
to explain the direct creation of our universe. 
These wonders are easier to comprehend in terms of
evolutionary theory - life started out in the shape
of some very simple molecules, which then inevitably evolved 
into the present forms of life. Similarly, a small patch of the
universe that satisfied some suitable properties underwent
inflation and inevitably 
evolved into our present universe. So Guth makes the correspondence
shown in Fig. 1.

\begin{figure}[tbp]
\centerline{
\epsfxsize = 0.8 \hsize 
\epsfbox{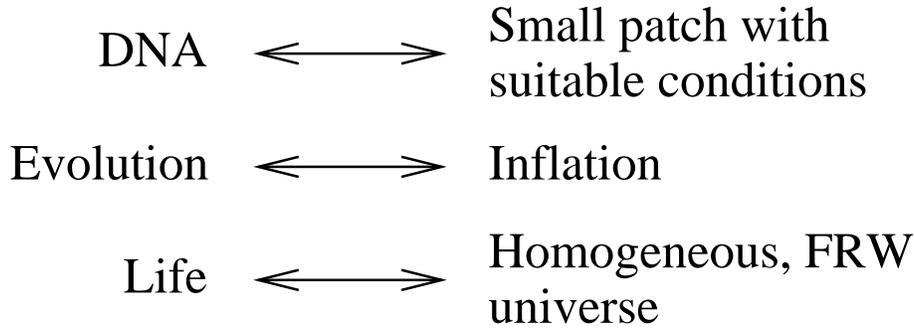}
}
\vskip 0.5 truecm
\caption{The correspondence proposed by Alan Guth.}
\end{figure}

I find this to be a very compelling argument for the existence of an
inflationary phase of the universe. 
The discussion in Ref. \cite{VacTro98}
impacts on this correspondence in the first stage - what is
the chance of getting a small patch with the correct conditions?
This is the same question that biologists may ask - what is the
chance of getting the first few molecules from which life can
follow? 

In the theory of evolution, the formation of the right kind of
molecules would depend on the geological and climatic conditions
at the time. For example, the occurrence of lightening storms 
could facilitate chemical reactions that could enhance the probability
of forming the molecules. Similarly we have found that the chance
of getting a small patch in the universe with all the
right conditions for inflation is very low, but that
there are conditions under which this probability can be
enhanced. So I would like to propose a further extension to
the correspondence as shown in Fig. 2.

Our philosophy in adopting inflation as a paradigm is that 
it greatly enhances the probability for the creation of the
universe that we see, even though it is at the expense of
invoking a fundamental scalar field - the inflaton. Then, since the 
introduction of the other factors in the extended correspondence can
further enhance the chances of creating an inflating universe,
this same philosophy guides us to include them as part
of the paradigm.

I should add that the preceding argument is not completely
obvious, as Guth and Linde explain, since inflation washes out the 
initial conditions and a scheme to obtain more likely initial
conditions may only provide an infinitesimal increase in the
final probability of observing a homogeneous universe.
However, if we consider a physical theory in which direct
creation of the universe, sub-horizon inflation, as well as 
super-horizon inflation are all possible, and in which the 
inflationary phases occur with the same expansion rates, it seems 
clear that the most spatial volume will be produced by the 
sub-horizon inflation. This will be true even in the case when 
the inflation is ``eternal''.

At this conference Koffman has raised the practical question of how 
these developments should affect the way inflationary cosmology is 
studied. I would first of all suggest recognition of the fact that
current inflationary models do not ``solve'' the homogeneity problem 
but only alleviate it. To solve the problem, one must resort
to relaxing at least one of the common assumptions. 
Linde has been advocating the path of Planckian inflation for
a variety of reasons.
The path of quantum cosmology has also been taken by a number of
researchers but, based on what I discussed above, these approaches 
appear to leave open a number of difficult questions that
need to be answered before we can claim to have understood the
homogeneity of the universe. To me two other approaches seem more 
promising. The first is to study the conditions under which
quantum effects can give rise to violations of the weak energy
condition (such as pursued by Ford and collaborators \cite{For}),
and subsequently to creation of baby universes 
(for an early related paper see \cite{FarGutGuv}). 
The second approach is to
explore modifications of the classical Einstein equations. 
This also ties in with the fact that essentially any high 
energy theory that is being considered these days does include 
such modifications. 

\begin{figure}[tbp]
\centerline{
\epsfxsize = 0.92\hsize 
\epsfbox{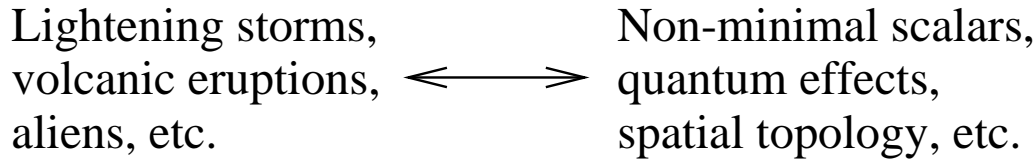}
}
\vskip 0.5 truecm
\caption{A proposed extension to Guth's correspondence.}
\end{figure}

If quantum effects or extensions to Einstein's
equations lead to the conclusion that inflation can begin from a
microphysical patch, profound consequences will follow since it
would open the possibility of creating universes in the laboratory. 
One could further imagine that stellar collapse, for example,
may lead to baby universe creation, and that there may 
be other universes lurking in the centers of galaxies.

It is a pleasure to thank Alan Guth, Andrei Linde and Alex Vilenkin 
for describing their views on inflation and also for their 
patience in this matter. I am grateful to Lawrence Krauss and
Mark Trodden for discussions. This work was supported by the 
Department of Energy.

\end{document}